\documentclass[twoside,preprint,english]{revtex4}
\usepackage[T1]{fontenc}

\makeatletter

\usepackage{babel}
\makeatother
\begin{document}

\title{The Homogeneous Scalar Field and the Wet Dark Sides of the Universe}

\author{Alberto Díez-Tejedor%
\footnote{wtbditea@lg.ehu.es%
} and Alexander Feinstein%
\footnote{a.feinstein@ehu.es%
}}

\address{Dpto. de Física Teórica, Universidad del País Vasco, Apdo. 644, 48080,
Bilbao, Spain.}

\begin{abstract}
We study the possibility that a generalised real scalar field minimally
coupled to gravity could explain \emph{both} the galactic and the
cosmological dark components of the universe. Within the framework
of Einstein's Relativity we model static galactic halos by considering
the most general action built from the scalar field and its first
derivatives. Although the gravitational configuration is static, the
scalar field may be either static, or homogeneous and linear in time.
In the case of the static scalar field, the models we look at inevitably
posses unphysical negative energies, and we are led to a sort of \emph{no-go}
result. In the case of the \emph{homogeneous} scalar field, on
the contrary, we find that compact objects with flat rotational curves
and with the mass and the size of a typical galaxy can be successfully
modeled and the Tully-Fisher relation recovered. We further show that the homogeneous scalar field deduced
from the galactic halo spacetimes has an action compatible with the
kinetic Unified Dark Matter models recently proposed by Scherrer.
Therefore, such a homogeneous kinetic Unified Dark Matter not only
may correctly mimic the galactic dynamics, but could also be used
to model the present-day accelerated expansion in the universe.
\end{abstract}

\pacs{95.35.+d, 95.36.+x, 98.62.Gq}

\maketitle

\section{Introduction}

Scalar fields play a central role in modeling the universe. Being
an essential ingredient in the inflationary scenario \cite{Guth},
the scalar fields might drive the initial accelerated expansion \cite{Linde}
necessary to solve some of the problems of the Cosmological Standard
Model. To explain the recent observations of type Ia supernovae \cite{Riess},
supported by other independent observations (Cosmic Microwave Background
(CMB) fluctuations, galaxy clustering), which suggest that we live
in an accelerated universe dominated by some kind of unknown Dark
Energy (DE) \cite{darkenergy}, scalar fields are again often evoked.
One can basically classify the scalar fields studied in the literature
into two main classes: the canonical ones \cite{Linde,canonical},
where the action is given by the sum of the standard kinetic and a
potential term, and the more general actions which were recently introduced
under the colloquial name of K-essence \cite{k-field}. It is worthwhile mentioning
that the non-canonical Lagrangians for the scalar field were studied earlier by Bekenstein 
and Milgrom \cite{bekenstein} in connection with the Modified Newtonian Dynamics. They also appear 
naturally in the velocity potential formulation of the Relativistic Hydrodynamics \cite{Schutz}.

One of the most challenging problems in theoretical cosmology and
astrophysics is the so-called Dark Matter (DM) problem (see \cite{DM,Sahni}
for a review). Several observations, carried out at different scales,
\emph{suggest} the existence of an invisible component, whose presence
can be only inferred through the gravitational effects. The main evidence
for the existence of DM appears at galactic scales. The mass profile
of a galaxy can be obtained by analyzing the rotation curves of the
surrounding particles-stars. This was done in the early 1980s \cite{Rubin}
by studying the velocity, obtained through the frequency shifts in
the 21 cm emission line of neutral hydrogen clouds. If we consider
that the circular orbits are stable due to the balance of the centrifugal
and gravitational forces, the mass profile in a galaxy must be given
by $M(r)=rv_{c}^{2}(r)/G$. It is observed that the velocity of rotation
is approximately an increasing function of the distance till the edge
of the visible galaxy, and remains nearly constant from there on.
The existence of some unobserved matter with an energy density profile
$\rho\propto r^{-2}$ is then inferred.

A different sign for the existence of DM appears at the scales
of galactic clusters. In 1933, Zwicky \cite{Zwicky} realized that
considering the clusters as equilibrium systems, their age being much
longer than their dynamical time-scale, the virial theorem $2\left\langle T\right\rangle +\left\langle U\right\rangle =0$
applies and the mass of the cluster can be evaluated with the result:
$M=\alpha\left\langle v^{2}\right\rangle R/G$. Here $\alpha$ is
a constant of order unity which depends on the matter distribution,
$M$ is the mass of the cluster and $R$ its radius. By determining
the speed of motion of many galaxies within a cluster from the shifts
of the spectral lines, Zwicky could infer the mass required to maintain
the Coma cluster held together. Surprisingly, it was by two orders
of magnitude greater than the observed one. Also the study of gravitational
lenses in galaxies \cite{galactic} and galaxy clusters \cite{cluster}
suggests that there should be more matter than the one obtained by
only counting stars.

Yet more evidence for the existence of a DM component comes from cosmological
scales. Recent data from the CMB \cite{CMB} point out that we live
in a flat ($k=0$), or a nearly flat universe \cite{flat}. This agrees
with the predictions of the inflationary scenario implying that one
must have $\Omega\simeq1$. The energy density of the universe, therefore,
must be very close to the critical value $\rho_{c}\sim10^{-29}$ g/cm$^{3}$,
pointing toward the discrepancy with the observed density in the universe
and the need of DM. Apart form this, DM seems to be an essential ingredient
in producing the observed structures in the Universe \cite{Mukhanov},
providing the gravitational potential depressions where the galaxies
are nourished and formed.

Many candidates for the DM component have been proposed along the
years \cite{DM}. They range from elementary particles such as WIMPs
(neutrinos, axions, neutralino, gravitino, etc.) to the ones in which
the DM is formed by compact objects such as primordial black holes
or MACHOs (brown dwarfs or Jupiter like objects). If one thing is
clear, however, is that the DM can not consist of baryonic matter
alone. The nucleosynthesis and the abundances of light elements \cite{nucleo}
restrict the amount of baryonic matter to $\rho_{b}\sim0.01\rho_{c}$,
therefore there must be something more out there to produce the observed
spatially flat universe. A sensible model for DM should probably solve
all the problems stated above at the same time, yet no model proposed
so far does the job.

Recently some suggestions to look at classical scalar fields as candidates
for the cosmological DM, rather than concentrating on fundamental
particles or some compact objects, were voiced \cite{Goodman}. Interesting models
have been proposed in which a scalar field, apart from modeling the
cosmological DM, can evolve to produce the accelerated expansion of
the universe. These models in which both cosmological dark components
(matter and energy) arise from the same field are commonly known as
Unified Dark Matter models. Some of the proposals involve the so-called
Chaplygin gas \cite{Chaplygin} and its generalisations \cite{Gen-Chap}.
In these models the universe expansion is driven by a perfect fluid
with a phenomenological equation of state derived from a string theory
motivated Born-Infeld Lagrangian \cite{sen}. The evolution of the
cosmological background interpolates between a DM universe at early
times and a DE dominated universe at the late epoch. Recent studies
indicate, however, the existence of serious problems related with
structure formation in these universes \cite{Problems} (see however \cite{however}). These problems
appear mainly due to a high speed of sound for the cosmological perturbations during certain periods
of the expansion, but may be alleviated in the so-called kinetic Unified
Dark Matter (kUDM) models recently introduced by Scherrer \cite{Scherrer},
where a speed of sound verifying $c_{s}^{2}\ll1$ is guaranteed for
all times. The actions in Scherrer's models depend only on the scalar
field derivatives maintaining the shift symmetry of the scalar field,
but otherwise are quite generic. Possible physical motivations for
these actions are discussed in \cite{k-field,arkani}. 

The main purpose of this paper is to analyse whether one could make
a step further and model in a consistent way, not only the cosmological
DM and DE, but also the DM one believes accounts for the observed
flat rotation curves in spiral galaxies, with a \emph{single} classical
scalar field. We consider spherically symmetric static galactic halos and
find that the generalised static scalar field configurations are not good candidates
to model the galactic matter. The homogeneous scalar field, on the
other hand, and one must admit somewhat surprisingly, leads to quite
interesting results. Not only the fact that the scalar field is homogeneous,
but rather the form of the action derived within the galactic halo,
akin to that one proposed by Scherrer \cite{Scherrer}, indicates that such matter
might be a perfect match for cosmology.

The paper is organised as follows. In Section II (and in the short
Appendix) we briefly discuss the form of the line element appropriate
for the problem of galactic halos. Section III deals with the matter
field. Here we show that if the gradient of the scalar field is spacelike,
the static spherically symmetric configurations of this field with
the imposed flattened rotational curves are unphysical. This goes
both for the canonical Quintessence and the non-canonical K-essence
fields, indicating a \emph{no-go} result. If, however, the gradient
of the field is timelike, the field is homogeneous and linear in time,
and the action for the scalar field contains only derivative terms
but not the field itself, one can get physically reasonable configurations.
The possibility of modeling the galactic halos with a homogeneous
scalar field opens a way to consider this field as the one also responsible
for the overall expansion of the universe. We find out that the model
of scalar matter suggested by the dynamics of the galactic halos fits
quite well with some simple unifying models suggested and studied
in cosmology by various authors under different names: x-matter \cite{x-matter},
\emph{wet} dark fluid \cite{wdm}, or the matter with the so-called
generalized linear equation of state \cite{babichev} (see as well \cite{Odintsov}). We study the
fitting of the free parameters of the action with galactic observations
and its implications for cosmology in the Section IV. Finally, in
the Section V, we discuss our results.

Throughout this paper we use the units in which $\hbar=c=1$, $M_{pl}=\left(8\pi G\right)^{-1/2}$
and the signature of the metric is taken to be $\left(-,+,+,+\right)$.

\section{Galactic Halos\label{sec:Galactic-Halos.}}

We consider that a typical galaxy is formed by a thin disk of visible
matter immersed in a large halo built of some unseen \emph{exotic}
matter, which can be conveniently described as static and spherically
symmetric. This exotic matter would be the main contributor to the
dynamics, so that the observed luminous matter can be treated as a
test fluid from which information about the physics of the halo can
be inferred. The observations indicate that the visible particles within the thin disk 
have rotation curves with velocities independent of their distance
to the centre of the galaxy \cite{Rubin,Persic}.

One wants to know how much information about the halo can be figured
out from the observed rotation curves. The Newtonian analysis mentioned
in the previous section is only valid under certain conditions \cite{Visser}:
weak gravitational field, small velocities and small pressures. In
this approximation, the gravitational field is given by the Newtonian
potential, which can be completely determined from the observed rotation
curves. The relativistic case is somewhat more complex.

The spherically symmetric and static halo in General Relativity is
described by a line element conveniently parametrized in \emph{curvature
coordinates} as \cite{wheeler}:\begin{equation}
ds^{2}=-e^{2\phi(r)}dt^{2}+\frac{1}{1-\frac{2m(r)}{r}}dr^{2}+r^{2}d\Omega^{2}.\label{eq:spher}\end{equation}
 In the equation above $d\Omega^{2}$ is the metric of a unit sphere,
and there are two metric functions: $\phi(r)$, known as the \emph{gravitational
potential}, and $m(r)$, known as the \emph{effective gravitational
mass}. In the Newtonian limit these two functions
co-incide with their usual interpretations. We distinguish the dynamical
mass \cite{lake} (or \emph{pseudo-mass} \cite{Visser}) $M(r)$ described
in the Introduction and obtained from the observed rotation curves,
and the effective gravitational mass $m(r)$, defined with the help
of the $rr$ component of the metric tensor and given in the expression
(\ref{eq:spher}). It is worthwhile to keep in mind that these are
two \emph{different} concepts and in general take different values. 

The most general static spherically symmetric metric with flat rotation
curves is given by (see the Appendix and the references \cite{lake,kar,sudarsky,matos}):\begin{equation}
ds^{2}=-\left(\frac{r}{r_{\star}}\right)^{l}dt^{2}+A(r)dr^{2}+r^{2}d\Omega^{2}.\label{eq:flat}\end{equation}
 Here $r_{\star}$ is a constant parameter with dimensions of length
and $l=2\left(v_{c}/c\right)^{2}$ . Therefore, the domain of the parameter
$l$ is restricted to $0<l<2$. To determine the metric function $A(r)$,
however, one needs to know more about the matter content. It is interesting
to point out that the form of the line element (\ref{eq:flat}), as
it stands, is generic in the sense that it does not depend on the
metric theory of gravity used, nor it depends on the matter content.
To obtain it one only assumes that the rotation curves are flat. In
short, the profile of the rotation curves gives us the chance of reproducing
completely the $00$ component of the metric tensor - the function
$\phi\left(r\right)$, but tells nothing about the other independent
component - the function $m(r)$ \cite{Visser,lake,kar,sudarsky,matos}.
If we want to obtain some information about the function $m(r)$,
one must assume either the nature of the matter dominating the configuration,
or deduce it from different observations, for example, gravitational
lensing \cite{Visser} etc.

In the observed galaxies the orbiting particles are non-relativistic,
therefore $l$ is a small parameter close to zero ($l<10^{-5}$).
The gravitational field far away from the centre, where super-massive
black holes are expected to ``hide'', is also small ($2m\left(r\right)/r$$\ll1$
and $2\phi\left(r\right)\ll1$) \cite{schodel}. The pressure inside
the halo, in the standard models of galaxies, is also usually assumed
to be close to zero ($p\ll\rho$). If these three conditions are met,
the Newtonian approximation applies and the parameters $m\left(r\right)$
and $M\left(r\right)$ do co-incide. Nevertheless, depending on the
choice of matter the last of the three above assumptions does not
always correspond to physical reality. Matter based on {}``string
fluids'' \cite{soleng} or on scalar fields \cite{sudarsky,matos,gonzalez,Armendariz}
could exert a significant amount of pressure and the Newtonian approximation
would no longer apply. An expression for the dynamical mass in a first
post-Newtonian approximation, which shows explicitly the discrepancies
between $M(r)$ and $m(r)$, is given in \cite{Visser}.

\section{The Dark Matter component}

To proceed any further one should specify the matter which makes up
the galactic halo. To model the DM we use the most general action
for a minimally coupled scalar field constructed from the scalar field
and its first derivatives:\begin{equation}
S=\int d^{4}x\sqrt{-g}\mathcal{L}(\varphi,X).\label{eq:action}\end{equation}
 Here $\mathcal{L}\left(\varphi,X\right)$ is the Lagrangian density
and $X$ the kinetic scalar defined by $X\equiv-1/2\: g^{\mu\nu}\partial_{\mu}\varphi\partial_{\nu}\varphi$.
These kind of actions are usually referred to as K-field, generalized
scalar field or, in a cosmological setting, K-essence \cite{k-field}.
Special cases discussed in the literature are the factorisable K-field
$\mathcal{L}=K(\varphi)F(X)$ \cite{factori} and the purely kinetic
K-field $\mathcal{L}=F(X)$ \cite{Scherrer}. The canonical scalar
field $\mathcal{L}=X-V\left(\phi\right)$ can be always rewritten
as a factorisable K-field by a field redefinition.

The energy-momentum tensor for the action (\ref{eq:action}) is:\[
T_{\mu\nu}=\frac{-2}{\sqrt{-g}}\frac{\delta S}{\delta g^{\mu\nu}}=\frac{\partial\mathcal{L}}{\partial X}\partial_{\mu}\varphi\partial_{\nu}\varphi+\mathcal{L}g_{\mu\nu}.\]
Here one runs into two \emph{qualitatively} different situations \cite{Taub}.
If the derivative term $\partial_{\mu}\varphi$ is timelike ($X>0$),
as it is usual in cosmology, we can identify the energy-momentum tensor
with a perfect fluid $T_{\mu\nu}=\left(\rho+p\right)u_{\mu}u_{\nu}+pg_{\mu\nu}$
by the formal relations:\begin{equation}
u_{\mu}=\frac{\partial_{\mu}\varphi}{\sqrt{2X}},\quad p=\mathcal{L},\quad\rho=2X\frac{\partial\mathcal{L}}{\partial X}-\mathcal{L}.\label{eq:isotro}\end{equation}
 Here $u_{\mu}$ is the 4-velocity, $\rho$ the energy density and
$p$ the pressure. On the other hand, if the derivative term is spacelike
($X<0$), this identification is no longer attainable. However, we
can still identify the energy-momentum tensor with an anisotropic
fluid $T_{\mu\nu}=\left(\rho+p_{\bot}\right)u_{\mu}u_{\nu}+p_{\bot}g_{\mu\nu}+\left(p_{\parallel}-p_{\bot}\right)n_{\mu}n_{\nu}$.
Now, the formal relations are:\begin{equation}
n_{\mu}=\frac{\partial_{\mu}\varphi}{\sqrt{-2X}},\quad p_{\bot}=-\rho=\mathcal{L},\quad p_{\parallel}=\mathcal{L}-2X\frac{\partial\mathcal{L}}{\partial X}.\label{eq:aniso}\end{equation}
 There are two different pressures in this case, one in the direction
parallel to $n_{\mu}$ ($p_{\parallel}$) and the other one in an
orthogonal direction ($p_{\bot}$). It is important to stress, however,
that the equality $p_{\bot}=-\rho$ for this sort of matter always
holds, implying that the Newtonian approximation is no longer valid.

The metric and the energy-momentum tensors are related through the
Einstein Equations $G_{\mu\nu}=M_{pl}^{-2}T_{\mu\nu}$, and since
we are only interested here in static, spherically symmetric configurations
(\ref{eq:spher}), restrictions on the function $\varphi\left(t,r\right)$
and on the form of the action $\mathcal{L}\left(\varphi,X\right)$
appear. The symmetries imply a diagonal Einstein and therefore a diagonal
energy momentum tensor, forcing the vanishing of the $T_{tr}$ component.
Therefore, either the scalar field is strictly static $\varphi=\varphi\left(r\right)$,
or it is strictly homogeneous $\varphi=\varphi\left(t\right)$. The inhomogeneous time dependent
scalar fields are not allowed by the symmetries imposed on the halo.  
Furthermore, the staticity of the spacetime ensures that the energy
density and the pressure must be independent of time. This implies
that for the case $\varphi=\varphi\left(t\right)$ the action can
only depend on the field derivatives $\mathcal{L}\left(\varphi,X\right)=F(X)$,
but not on the field itself. Moreover the scalar field $\varphi$
may only be linear in time, $\varphi=at$. It is interesting to observe
that in the last case the spacetime does not inherit the symmetries
of the fundamental field producing the geometry, and while the metric
remains static, the fundamental scalar field may be time dependent.
Similar, but rather an inverse situation was discussed in the case
of a tilted homogeneous string cosmology \cite{Clancy}, where the
homogeneous geometry was produced by an inhomogeneous scalar dilaton.

\subsection{The static field $\varphi=\varphi\left(r\right)$}

In this case the derivative of the field is spacelike ($X<0$) and
the scalar field, therefore, behaves as an anisotropic fluid (\ref{eq:aniso}).
The Einstein equations for the $tt$, $rr$ and $\theta\theta$ components
are:\begin{equation}
\frac{1}{Ar^{2}}\left[\frac{A'}{A}r+A-1\right]=-M_{pl}^{-2}\mathcal{L},\label{eq:tt1}\end{equation}
\begin{equation}
\frac{1}{Ar^{2}}\left[\left(l+1\right)-A\right]=-M_{pl}^{-2}\left[2X\frac{\partial\mathcal{L}}{\partial X}-\mathcal{L}\right],\label{eq:rr1}\end{equation}
\begin{equation}
\frac{1}{4Ar^{2}}\left[\left(l+2\right)\frac{A'}{A}r-l^{2}\right]=-M_{pl}^{-2}\mathcal{L}.\label{eq:331}\end{equation}
 Here the prime denotes differentiation with respect to the variable
$r$. The above three equations imply automatically the matter conservation
equation. Combining (\ref{eq:tt1}) and (\ref{eq:331}) we obtain
a differential equation for the metric component $A(r)$:\[
\left(l-2\right)\frac{A'}{A}r-4A-\left(l^{2}-4\right)=0,\]
 whose general solution is given by: \begin{equation}
A\left(r\right)=\frac{a_{i}}{1\pm\left(\frac{R_{i}}{r}\right)^{b_{i}}}.\label{eq:sol}\end{equation}
 Here $R_{i}$ is a positive arbitrary integration constant with dimensions
of length. We have introduced the subindex $i$ to distinguish between
the cases $\varphi=\varphi\left(r\right)$ ($i=r$) and $\varphi=\varphi\left(t\right)$
($i=t$) of the following subsection. The strictly positive parameters
$a_{r}$ and $b_{r}$ are given by:\[
a_{r}=\frac{4-l^{2}}{4},\quad b_{r}=2+l.\]
 Simple analytic solutions can be easily obtained for the special
case $R_{r}=0$. Such are the solutions for the canonical scalar field
with an exponential potential \cite{matos}, the factorisable K-field
with a potential of the form $K(\varphi)=1/\varphi^{2}$ and arbitrary
$F(X)$ \cite{unpu} or the purely kinetic case with $F(X)\propto X^{-2/l}$
\cite{unpu}. We do not give the explicit solutions here due to their
scarce physical relevance. For all these solutions the metric function
$A(r)$ is a constant and the energy density and the pressure are
obtained directly from the equations (\ref{eq:tt1}) and (\ref{eq:rr1})
using the definitions given in (\ref{eq:aniso}):\[
p_{\bot}=-\rho=\frac{l^{2}}{4-l^{2}}\left(\frac{M_{pl}}{r}\right)^{2},\quad p_{\parallel}=\frac{l\left(l+4\right)}{4-l^{2}}\left(\frac{M_{pl}}{r}\right)^{2}.\]
 It is easy to see that the energy density is always negative, therefore
these solutions must be considered unphysical. This affects as well
the solution presented in \cite{matos}, where the energy density
sign (their expression (19)) is erroneous. Note that the energy density
and the pressure are directly read from the line element and as long
as the galactic rotational velocities are non-tachyonic one can not
escape the negative energy densities.

For the general case ($R_{r}\neq0$) the situation is more complex
and exact analytic expressions for the action can not be easily given.
However, certain observations about the general behavior of the solution
can still be made. In the asymptotic limit $r\gg R_{r}$ the general
solution converges to the special case $R_{r}=0$ mentioned above,
so the negative energy densities persist. These solutions, nevertheless,
could still be useful if their behavior were physical for some range
$r<r_{crit}$ say, and then matched to a different solution. However,
unfortunately, this is not the case. For the $-$ branch of the equation
(\ref{eq:sol}) the energy density remains always negative. For the
$+$ branch, positive energy densities appear for some range $r<r_{crit}$,
but the effective gravitational mass $m(r)$ defined by (\ref{eq:spher})
remains always negative, so the solution is again unphysical. The occurrence
of the negative mass with a positive energy density signals the presence
of a singularity with an infinite negative mass in the centre, analogous
to the one discussed by Bonnor in \cite{bonnor}.

Thus, one may consider this subsection as one leading to a kind of
\emph{no-go} result for the static scalar fields $\varphi=\varphi\left(r\right)$.
Once we have assumed the form of the line element suitable for the
description of flat rotation curves (Section \ref{sec:Galactic-Halos.}),
the minimally coupled generalised \emph{static} scalar field is found
unfit to play the role of the dark matter in the galactic halos.

\subsection{The homogeneous field $\varphi=at$}

The configuration $\varphi=at$, as mentioned above, is only possible
if the action (\ref{eq:action}) is of the purely kinetic form $\mathcal{L}\left(\varphi,X\right)=F(X)$
- no potential term. In this case, one may interpret the scalar field
as an irrotational isentropic perfect fluid in a disguise \cite{nosotros},
although the equation of state $p=p(\rho)$ does not have to be of
a simple form. The action itself can be thought of as the hydrodynamical
action written in terms of the velocity potential \cite{nosotros,Schutz}. 

The $tt$, $rr$ and $\theta\theta$ components of the Einstein equations
are now:

\begin{equation}
\frac{1}{Ar^{2}}\left[\frac{A'}{A}r+A-1\right]=M_{pl}^{-2}\left[2X\frac{\partial\mathcal{L}}{\partial X}-\mathcal{L}\right],\label{eq:tt2}\end{equation}
\begin{equation}
\frac{1}{Ar^{2}}\left[\left(l+1\right)-A\right]=M_{pl}^{-2}\mathcal{L},\label{eq:rr2}\end{equation}
\begin{equation}
\frac{1}{4Ar^{2}}\left[-\left(l+2\right)\frac{A'}{A}r+l^{2}\right]=M_{pl}^{-2}\mathcal{L}.\label{eq:332}\end{equation}
 Combining the expresions (\ref{eq:rr2}) and (\ref{eq:332}), we
obtain a differential equation for the function $A(r)$:

\[
\left(l+2\right)\frac{A'}{A}r-4A-\left[l^{2}-4\left(l+1\right)\right]=0,\]
 whose general solution is again given by (\ref{eq:sol}), where the
parameters $a_{t}$ and $b_{t}$ are now:\[
a_{t}=-\frac{l^{2}-4\left(l+1\right)}{4},\quad b_{t}=\frac{l^{2}-4\left(l+1\right)}{l+2}.\]
 Unlike in the static case, however, while the parameter $a_{t}$
still remains positive, the parameter $b_{t}$ becomes negative. 

For the special case when the constant $R_{t}=\infty$ an analytic
solution can be easily found. The action is given by $F(X)\propto X^{2/l}$,
and the energy density and the pressure are:\[
\rho=\frac{l\left(l-4\right)}{l^{2}-4\left(l+1\right)}\left(\frac{M_{pl}}{r}\right)^{2},\quad p=\frac{-l^{2}}{l^{2}-4\left(l+1\right)}\left(\frac{M_{pl}}{r}\right)^{2},\]
 which are both positive. Interpreted as a fluid with a constant barotropic
index $p=w\rho$ one gets $w=l/(4-l)$ ($0<w<1$), so that for the
observed galaxies ($l\ll1$) the fluid is nearly {}``dust''. This
solution is analogous to the infinite \emph{isothermal sphere.}

In the general case ($R_{t}\neq\infty$) the solution is rather more
interesting. We will be interested in the $-$ branch of the equation
(\ref{eq:sol}). In this case, the complete solution can be derived
from a scalar field obeying the following two-parameter family of
K-actions:\begin{equation}
F(X)=\left(\frac{M_{pl}}{R_{t}}\right)^{2}\left[p_{1}^{N}p_{2}^{2}\right]^{\frac{1}{N+2}}\left[X^{2/l}-X^{-N/l}\right].\label{eq:exactF}\end{equation}
For these fluids, the energy density and the pressure are:\[
\rho=\rho_{1}\left(\frac{M_{pl}}{r}\right)^{2}+\rho_{2}\left(\frac{M_{pl}}{R_{t}}\right)^{2}\left(\frac{r}{R_{t}}\right)^{N},\]
 \begin{equation}
p=p_{1}\left(\frac{M_{pl}}{r}\right)^{2}-p_{2}\left(\frac{M_{pl}}{R_{t}}\right)^{2}\left(\frac{r}{R_{t}}\right)^{N},\label{eq:pressure}\end{equation}
 where $\rho_{1}$, $\rho_{2}$, $p_{1}$, $p_{2}$ and $N$ are positive
constants given by:\[
\rho_{1}=\frac{-l\left(4-l\right)}{l^{2}-4\left(l+1\right)},\quad\rho_{2}=\frac{-4\left(6+5l-l^{2}\right)}{\left(l+2\right)\left[l^{2}-4\left(l+1\right)\right]},\]
 \[
p_{1}=\frac{-l^{2}}{l^{2}-4\left(l+1\right)},\quad p_{2}=\frac{-4\left(l+1\right)}{l^{2}-4\left(l+1\right)},\]
 \[
N=\frac{l\left(2-l\right)}{l+2}.\]
 The solutions are only valid for $r<R_{t}$ due to the behavior of
the metric component given by the Eq.(\ref{eq:sol}), and are in fact
the Tolman type V solutions \cite{Tolman}. As pointed out by Tolman,
and as is usual for solutions of this kind, an explicit equation
of state $p=p\left(\rho\right)$ can not be found. It is interesting,
however, to point out that it has been possible to obtain a simple
action which describes the fluid (\ref{eq:exactF}) without referring
to an explicit equation of state. In fact, a simple action can be
often given in many instances for which the explicit equation of state
$p\left(\rho\right)$ is not available.

Further, these solutions may be used to describe compact objects of
finite size. The expression (\ref{eq:pressure}) indicates that the
pressure is positive until some value $r_{0}$ is reached, then it
vanishes and changes sign. It is thus possible to match the solution
with an exterior Schwarzschild vacuum for $r>r_{0}$, defining $r=r_{0}$
as the halo external surface. We will use the subindex $0$ to refer
to the values evaluated on this surface. The expressions for the radius
$r_{0}$ and for the effective gravitational mass $m_{0}=4\pi\int_{0}^{r_{0}}\rho\left(r\right)r^{2}dr$
of the compact objects are then given (in physical units) by:\[
r_{0}=\left(\frac{p_{1}}{p_{2}}\right)^{\frac{1}{2+N}}R_{t},\quad m_{0}=\frac{l}{2(l+1)}\frac{c^{2}r_{0}}{G}.\]
The constant $r_{\star}$ of the equation (\ref{eq:flat}) is now fixed by the matching to 
the exterior vacuum solution and is given by $r^l_{\star}=(l+1)r^l_0$. 
Since $r_{0}<R_{t}$, it is always possible to construct these solutions.
We are only interested in the behavior of the fluid within the halo,
which in turn restricts the domain of the function $F(X)$ to $X\geq1.$ 

The expressions above are \emph{exact}. However, we are more interested
in a \emph{rule of thumb} to work with, and since the observed rotation
velocities in the galaxies are small compared to the speed of light,
we have found it convenient to proceed working to first  
order in $l$. To this order we propose the
simplified action\begin{equation}
F(X)=\left(\frac{M_{pl}}{R_{t}}\right)^{2}\left[X^{2/l}-1\right],\label{eq:F(X)}\end{equation}
which fits amazingly well with (\ref{eq:exactF}) within range $X\geq1$, where
the equation (\ref{eq:exactF}) applies. The result above is central to this paper. It
gives a simple equation of state, or rather an action for the homogeneous
scalar field, which can be used to model the DM in the galaxies. The 
equation (\ref{eq:F(X)}) should not be seen as a formal limit of the equation
(\ref{eq:exactF}), but rather as a suggested, or guessed action describing 
the matter. This matter approximates the equation (\ref{eq:exactF}) within the
halo, and therefore reproduces the desired geometry. Moreover,
similar actions are used in cosmology under the different names of
x-matter \cite{x-matter}, wet dark fluid \cite{wdm} and matter with
the generalized linear equation of state \cite{babichev}. Thus, the action (\ref{eq:F(X)})
may potentially serve as a unified matter description both for cosmology, on
one hand, and within the galactic halo, on the other. The two
arbitrary constants in the model: $R_{t}$ and $l$, must be determined
from the observations. A brief estimate of the order of magnitude
of these parameters will be the task of the following section.

\section{Fitting the model}

\subsection{Galaxies}

We first adjust the two free parameters of the equation of state to
fit a typical galaxy. As we have mentioned in the Section \ref{sec:Galactic-Halos.},
the parameter $l$ is directly related to the rotation velocities.
We take $l\sim10^{-5}$. The size and the mass of a typical galaxy
are then given by the first order expressions:

\[
r_{0}\simeq\frac{l}{2}R_{t},\quad m_{0}\simeq\frac{l}{2}\frac{c^{2}r_{0}}{G}.\]
If we take $R_{t}$ of the order $R_{t}\sim3\,000\:\textrm{Mpc}$
one obtains a compact object with $r_{0}\sim15\:\textrm{Kpc}$ and
$m_{0}\sim10^{12}M_{\odot}$, compatible with the size and the mass
for a typical galactic halo. Curiously enough we had to assume the
constant $R_{t}$ of the order of the size of the observable universe
to fit the observations. The last two equations can be combined to
obtain a relation between the mass of a galaxy and the velocity of
the orbiting particles:\[
m_{0}\simeq\frac{R_{t}}{c^{2}G}v_{c}^{4}.\]
The equation above is nothing else but the Tully-Fisher relation ($m_{0}=\kappa v_{c}^{4}$),
where the constant of proportionality $\kappa$ is determined by the
parameter $R_{t}$ of the model.

Now, the action we propose to describe the DM
in the halos of the spiral galaxies (\ref{eq:F(X)}) with the parameters
$R_{t}\sim10^{3}\:\textrm{Mpc}$ and $l\sim10^{-5}$ may work. But
then one may possibly suggest to identify the homogeneous
scalar field with the velocity potential of some perfect fluid in
the galaxy and just stop here, instead of further promoting
it to cosmic scales. If the equation of state we propose were related
exclusively to the galactic matter, it would seem unlikely, that the
same action parametrised by the same parameters could fit the global
universal expansion. The presence of the constant $R_{t}$ of the
order of the size of the observable universe in the parametrisation
of the galactic DM action signals as well a tentative relation to
cosmology. More interestingly, the action (\ref{eq:F(X)}) does fit
favorably with cosmology, and this is a non-trivial result.

\subsection{Cosmology}

Consider now the action (\ref{eq:F(X)}) in the setting of a homogeneous
and isotropic universe. Working still to first order in $l$, the
pressure and the energy density are easily obtained using the relations
given in (\ref{eq:isotro}):

\[
p=\left(\frac{M_{pl}}{R_{t}}\right)^{2}\left[X^{2/l}-1\right],\quad\rho\simeq\left(\frac{M_{pl}}{R_{t}}\right)^{2}\left[\frac{4}{l}X^{2/l}+1\right].\]
 For convenience it is useful to define the parameters $w$ and $c_{s}^{2}$,
the barotropic index and the velocity of sound respectively, given
by the following expressions :

\[
w=\frac{F(X)}{2XF'(X)-F(X)}\simeq\frac{X^{2/l}-1}{\frac{4}{l}X^{2/l}+1},\]
\begin{equation}
c_{s}^{2}=\frac{F'(X)}{2XF''(X)+F'(X)}\simeq\frac{l}{4}.\label{eq:sound}\end{equation}
 The barotropic index $w$ defines the evolution of the scale factor of such a universe,
while the sound speed $c_{s}^{2}$ gives the evolution of the first
order small perturbations \cite{Garriga}. There is no need to write
down the solutions of the Einstein Equations to see how the model
behaves. The scalar field evolution equation is given by:\[
\dot{X}+6Hc_{s}^{2}\left(X\right)X=0.\]
Here $H=\dot{a}/a$ is the Hubble constant, $a$ is the scale factor
and the dot has the usual meaning. While the universe expands ($H>0$)
and given that the square of the speed of sound (\ref{eq:sound})
remains always positive, the kinetic scalar $X$ is a decreasing function
of time. The barotropic index evolves from {}``dust'' ($w\simeq0$) at
early times ($X\gg1$), to a {}``cosmological constant'' ($w\simeq-1$)
at late times ($X\ll$1). The speed of sound, however, always remains
small, $c_{s}^{2}\ll1.$ Even during the late epoch, when the fluid
approaches the {}``cosmological constant'' like equation of state,
the sound speed remains well below the speed of light. The fact that
the DE density perturbations propagate with low velocities makes this
model distinguishable from the standard $\Lambda$CDM, where a purely
non-clustering Cosmological Constant would produce a different pattern
at large angular scales in the CMB fluctuations \cite{HU,erickson}.

The energy density for the effective value of the {}``cosmological
constant'' is determined by the parameter $R_{t}$ fixed in the previous
subsection against the galactic data and is given by \[
\rho_{\Lambda}=\frac{c^{2}}{8\pi GR_{t}^{2}}=6.5\cdot10^{-30}\textrm{g/cm}^{3}.\]
The value of the {}``cosmological constant'' then becomes $\Lambda_{eff}=R_{t}^{-2}$
which gives $\Lambda_{eff}\sim10^{-52}cm^{-2}$ (cf. \cite{darkenergy}). 

We also obtain that there exists a relation between the value of the
cosmological constant and the Tully-Fisher proportionality factor
$\kappa$:\[
\kappa^{2}\Lambda_{eff}=\frac{1}{G^{2}c^{4}},\]
determined only by the fundamental constants $G$ and $c$.

\section{Discussion}

The flattening of the rotation curves in spiral galaxies, the missing
mass in the galactic clusters, the spatial flatness of the universe,
the structure formation and the present day acceleration all point
out to the existence of a Dark Side in the Universe. It is possible
that each of the above mentioned problems has a separate solution;
it would be physically more appealing, though, if the solution were
unique. 

Our Universe is homogeneous and isotropic on large scales. On galactic
scales, however, the observed matter distribution is non-uniform.
This itself, does not exclude the possibility of the presence of a
homogeneous scalar field on scales of galactic halos in addition to
a non-uniform ordinary matter distribution. 

We have started the technical part of the paper by looking at possible
static configurations of scalar field compatible with flat rotational
curves, and have concluded that if the scalar field is static no physically
interesting solutions result. 

If the scalar field is homogeneous, however, one may construct models
of galactic halos based on exact solutions of the Einstein Equations
which have flat rotational curves. The matter in these galactic halos
is described by an action which only depends on derivative terms of
the scalar field. We interpret this intra-galactic scalar field as
an integral part of a global homogeneous field driving the expansion
of the universe. We do so, encouraged by the fact that the homogeneous
scalar field we deduce from the galactic rotational curves makes a
good match to cosmology. 

The model for the homogeneous field we have read from the galactic
halo metric is of the form which belongs to a class of models which
could solve the Dark Matter and the Dark Energy problem in cosmology
- the class of models suggested by Scherrer \cite{Scherrer}. Of course,
this could be a pure co-incidence: the same action for the homogeneous
scalar field resolves it all. Nonetheless, we believe it is worth exploring
this possibility further.

The most interesting feature of the model is, as we see it, that the
two free parameters in the K-action, when fixed against the galactic
data, imply a reasonable value for the cosmological constant which
is unrelated to the fit. There also exist the possibility to differentiate
the model from the standard $\Lambda$CDM due to a different clustering
of the Dark Energy in our model which would leave observable imprints
on the Cosmic Microwave Background.

Some words of caution should be spelled out. One should of course
take the global solution to the DM problem we suggest as a toy model.
The action for the cosmological fluid is given by the Eq. (\ref{eq:F(X)})
with the parameters fixed by the galactic curves. Now, while the constant
$R_{t}$ is universal in the model, different galaxies would have
different rotation velocities and therefore different values for $l$.
But in cosmology one must fix the value of $l$ presumably globally.
One may speculate that the global value of the constant $l$ in the
cosmological equation of state is determined as some sort of average,
or is given by the first principles of an underlying theory. Be what
may, however, it is encouraging to find out that there exists the
possibility that a \emph{single} matter ingredient can be used to
model the Dark Side of the universe on all scales. 

\begin{acknowledgments}
We are grateful to Jacob Bekenstein for correspondence and valuable comments.
A.D.T. work is supported by the Basque Government predoctoral fellowship
BFI03.134. This work is supported by the Spanish Science Ministry
Grant 1/MCYT 00172.310-15787/2004, and the University of the Basque
Country Grant 9/UPV00172.310-14456/2002. 
\end{acknowledgments}

\section*{APPENDIX}

\section*{Determination of the $g_{00}$ component of the metric tensor from
the rotation curves}

In this Appendix we briefly summarize how one can reconstruct some
metric functions from the observed rotation curves \cite{lake,kar,sudarsky,matos}.
We consider the equations of motion for an orbiting particle of mass
$m$ immersed in a static spherically symmetric gravitational field
(\ref{eq:spher}). For this purpose we choose, as usual, the coordinate
system so that the orbit is contained in the equatorial plane $\theta=\pi/2$.
The two conserved quantities, associated with the two Killing vectors
of the metric $\xi_{t}=\partial/\partial t$ and $\xi_{\phi}=\partial/\partial\phi$,
are respectively the energy $E\equiv-p\cdot\xi_{t}=-p_{0}$ and the
angular momentum $L\equiv p\cdot\xi_{\phi}=p_{\phi}$, where $p$
is the 4-momentum of the particle. The equation of motion for the
radial coordinate is obtained from the constraint $p\cdot p+m^{2}=0$
by introducing the values for $E$ and $L$ given above, and can be
written as:\[
f\left(r\right)\dot{r}^{2}=\tilde{E}^{2}-\tilde{V}_{eff}^{2}\left(r\right),\]
with $f\left(r\right)$ and $\tilde{V}_{eff}^{2}\left(r\right)$ given by:\[
f\left(r\right)=\frac{e^{2\phi\left(r\right)}}{1-\frac{2m\left(r\right)}{r}},\quad \tilde{V}_{eff}^{2}\left(r\right)=e^{2\phi\left(r\right)}\left(1+\frac{\tilde{L}^{2}}{r^{2}}\right).\]
The symbol $\sim$ indicates quantities evaluated per unit mass. 

The circular orbits are obtained by imposing $\dot{r}=0$ ($\tilde{V}_{eff}\left(r\right)=\tilde{E}$)
and $\ddot{r}=0$ ($\tilde{V}'_{eff}\left(r\right)=0$). This gives the following
expressions for $\tilde{E}$ and $\tilde{L}$:\[
\tilde{E}^{2}=\frac{e^{2\phi}}{1-r\phi'},\quad\tilde{L}^{2}=\frac{r^{3}\phi'}{1-r\phi'}.\]
Using the definition for the tangential component of the velocity
(equation (25.20) in the reference \cite{wheeler}) and the expressions
for $\tilde{E}$ and $\tilde{L}$ above, we obtain:\[
v_{c}^{2}=\frac{p_{\phi}^{2}}{r^{2}E_{local}^{2}}=r\phi'.\]
Here $v_{c}$ is measured in units of $c$ ($v_{c}\rightarrow v_{c}/c$)
and $E_{local}$ is the energy of the particle measured by a local
observer at rest. From the behavior of the rotation curves $v_{c}\left(r\right)$
the function $\phi\left(r\right)$ can be directly integrated and
the $00$ component of the metric tensor obtained:\begin{equation}
g_{00}\left(r\right)=-\exp\left(2\int\frac{v_{c}^{2}\left(r\right)}{r}dr\right).\label{eq:g00}\end{equation}
For a galaxy with flat rotation curves ($v_{c}(r)=const.$), one obtains:\begin{equation}
g_{00}\left(r\right)=-\left(\frac{r}{r_{\star}}\right)^{l},\label{eq:g00flat}\end{equation}
where $r_{\star}$ is an integration constant with dimensions of length
and $l$ is given by $l=2v_{c}^{2}$.

Two comments are in order. First, that the form of the rotation curve
completely determines the $00$ component of the metric tensor through
the expression (\ref{eq:g00}). For the special case of flat rotation
curves, we obtain (\ref{eq:g00flat}). Second, that the rotation curves
tell nothing about the other independent component of the metric tensor
(the $rr$ component). To obtain the $rr$ component of the line element,
one needs to assume either the nature of the matter dominating the
configuration, or deduce it from other type of observations, for example,
gravitational lensing \cite{Visser} etc.

\end{document}